\documentclass[aps,prc,twocolumn,floatfix,superscriptaddress,nofootinbib,preprintnumbers,longbibliography,amsmath,amsthm,amssymb]{revtex4-2}
\pdfoutput=1
\usepackage{graphicx}
\usepackage{color}
\usepackage[colorlinks=true,linkcolor=blue,citecolor=blue]{hyperref}
\usepackage{dcolumn}
\usepackage{bm}

\newcommand{\be}{\begin{equation}}
\newcommand{\ee}{\end{equation}}
\newcommand{\dd}{\mathrm{d}}
\newcommand{\br}{\boldsymbol{r}}

\begin{document}

\title{Isovector spin susceptibility: Isotopic evolution of collectivity in spin response
}

\author{Kenichi Yoshida}
\email[E-mail me at: ]{kyoshida@ruby.scphys.kyoto-u.ac.jp}
\affiliation{Department of Physics, Kyoto University, Kyoto, 606-8502, Japan}
\preprint{KUNS-2866}
\date{\today}

\begin{abstract}
\begin{description}
\item[Background] Response to spin-dependent operators 
has been investigated in the magnetic dipole and Gamow--Teller transitions, 
which provides magnetic properties of a nuclear system.
\item[Purpose] I investigate an isotopic dependence of the collectivity generated by the spin-dependent interactions 
in the Ca and Ni isotopes 
through the isovector (IV) spin-flip excitations. 
The responses in the neutral $(t_z)$ and charge-exchange $(t_{\pm})$ channels are 
considered in a unified way. 
\item[Method] A nuclear energy-density functional approach is employed for calculating the response functions 
based 
on the Skyrme--Kohn--Sham--Bogoliubov method and the quasiparticle-random-phase approximation (QRPA). 
I adopt the like-particle QRPA and the proton--neutron QRPA for the neutral 
and charge-exchange channels, respectively. I consider 
the fluctuation of the proton--neutron pair fields.
\item[Results] The collective shift due to RPA correlations for the response in the neutral channel 
is explained by the occupation probability of neutrons in the $j_>=\ell+1/2$ orbital. 
Many particle--hole or two-quasiparticle excitations have a coherent 
contribution to form a giant resonance in neutron-rich nuclei for the charge-exchange channel. 
The IV spin susceptibility displays the isotopic evolution of the collectivity and the underlying shell structure. 
\item[Conclusions] A repulsive character of the residual interaction in the spin--isospin channel diminishes 
the IV spin susceptibility due to the collectivity, 
while the dynamic ${}^{3}S$ pairing appearing in the charge-exchange channel opposes the reduction. 
\end{description}
\end{abstract}

\maketitle

\section{Introduction}\label{intro}
Susceptibility to an applied magnetic field is a fundamental property characterizing the matter. 
A detailed investigation unveils the single-particle and collective behavior of the spin degree of freedom.
An interplay of superconductivity and magnetism
has attracted much interest
and has been a central issue in strongly-correlated many-body systems~\cite{bul85}.
The fermionic superfluidity or superconductivity 
is well understood by the $^1S_0$ pair correlation mostly~\cite{sch64,bri05}.
The onset of the spin susceptibility below the critical temperature is a manifestation of the spin-triplet $(S=1)$
superfluidity of $^3$He~\cite{bal63,leg65}. 
The $S=1$ pairing in electronic systems has been investigated not only in the superfluid $^{3}$He~\cite{vol90} 
but in other systems recently~\cite{rob02,mac03}.
In contrast, there has been a continuing discussion on 
the $S=1$ pairing in nuclear systems: 
the $^3S_1$ correlation of isoscalar (IS) proton--neutron (pn) pairs in $N \sim Z$ nuclei~\cite{fra14} 
and the $^3P_2$ pairing of neutrons in the neutron-star matter~\cite{hae07}.

Response of a nucleus to an external field displays 
the correlations among constituent nucleons. 
The nuclear response is characterized by the transferred angular momentum 
$\Delta L$, spin $\Delta S$, isospin $\Delta T$, and nucleon number $\Delta N$~\cite{har01}.
The response categorized as $\Delta L=0, \Delta S=0, \Delta T=1,\Delta N=2$ 
pins down the $^1S_0$ pairing~\cite{bri05}, 
and the response $\Delta L=0, \Delta S=1, \Delta T=0, \Delta N=2$ 
has been studied recently to uncover the $^3S_1$ pairing~\cite{yos14,lit18,cha21}.
The response to the spin-dependent electromagnetic and weak probes 
reveals the magnetic properties of nuclear matter~\cite{eri84,ost92,hey10}. 
The IV spin magnetic-dipole (M1) response with $\Delta L=0, \Delta S=1, \Delta T=1, \Delta N=0$
has been investigated extensibly because they 
impact diverse quantities such as 
the neutral-current neutrino-nucleus cross sections relevant to 
supernova physics and neutrino physics~\cite{nak17} and 
the neutron-capture cross sections relevant to the nucleosynthesis~\cite{kap11,arn07,kaj19}. 
Furthermore, the IV spin-M1 response 
can be seen in a wider perspective 
when it is considered as a single component $\Delta T_z=0$ of the IV modes~\cite{aue84}. 
The additional components $\Delta T_z= \pm 1$ represent the charge-exchange modes, namely, 
the Gamow--Teller (GT) response. 

The GT response can be a probe to disclose the $^3S_1$ IS pairing. 
The dynamic IS pairing lowers the GT states in energy and thus shortens the 
$\beta$-decay half-lives of neutron-rich nuclei~\cite{eng99} including deformed nuclei~\cite{yos13}.
The effects of the IS pairing on the GT transition strengths have been studied in $N = Z$ 
odd-odd nuclei with a three-body model of two nucleons around a spherical core~\cite{tan14}. 
A remarkable feature found in $N=Z$ odd-odd nuclei with an $LS$-closed core ($^4$He, $^{16}$O, $^{40}$Ca) is 
the appearance of the low-energy state with a strong GT strength. 
A similar trait is also found by employing a microscopic nuclear energy-density functional (EDF) method~\cite{bai14}.
The low-energy GT states have been indeed identified experimentally 
in the transitions of $^{18}\text{O}\to{}^{18}\text{F}$~\cite{fuj19} 
and $^{42}\text{Ca}\to{}^{42}\text{Sc}$~\cite{fuj14,fuj15}.

In this article, I am going to investigate 
the collectivity generated by the spin response in view of the pair correlations. 
An isotopic dependence is studied to see the shell effects. 
To this end, the 
EDF method is employed: a theoretical model capable of 
handling nuclides with an arbitrary mass number in a single framework~\cite{ben03,nak16}. 
I introduce and evaluate the IV spin susceptibility to discuss the neutron number dependence
 of the magnetic property quantitatively.

This paper is organized in the following way: 
the theoretical framework for describing the spin responses is given in Sec.~\ref{method} and 
details of the numerical calculation are also given; 
Sec.~\ref{result} is devoted to the numerical results and discussion based on the model calculation; 
the discussion on the response of the Ca isotopes in the neutral and charge-exchange channels 
is given in Sec.~\ref{result_M1} and Sec.~\ref{result_GT}, respectively; 
the discussion for the Ni isotopes is given in Sec.~\ref{Ni_isotopes}; 
then, a summary is given in Sec.~\ref{summary}.

\section{Framework}\label{method}

\subsection{KSB and QRPA for neutron-rich nuclei}

Since the details of the formalism can be found in Refs.~\cite{yos08,yos13b,yos13}, 
here I briefly recapitulate the basic equations relevant to the present study. 
In the framework of the nuclear EDF method I employ, 
the ground state of a mother (target) nucleus is described by solving the 
Kohn--Sham--Bogoliubov (KSB) equation~\cite{dob84}:
\begin{align}
\sum_{s^\prime}
\begin{bmatrix}
h^q_{s s^\prime}(\br)-\lambda^{q}\delta_{s s^\prime} & \tilde{h}^q_{s s^\prime}(\br) \\
\tilde{h}^q_{s s^\prime}(\br) & -h^q_{s s^\prime}(\br)+\lambda^q\delta_{s s^\prime}
\end{bmatrix}
\begin{bmatrix}
\varphi^{q}_{1,\alpha}(\br s^\prime) \\
\varphi^{q}_{2,\alpha}(\br s^\prime)
\end{bmatrix} \notag \\
= E_{\alpha}
\begin{bmatrix}
\varphi^{q}_{1,\alpha}(\br s) \\
\varphi^{q}_{2,\alpha}(\br s)
\end{bmatrix}, \label{HFB_eq}
\end{align}
where 
the single-particle and pair Hamiltonians $h^q_{s s^\prime}(\br)$ and $\tilde{h}^q_{s s^\prime}(\br)$ 
are given by the functional derivative of the EDF with respect to the particle density and the pair density, respectively. 
An explicit expression of the Hamiltonians is found in the Appendix of Ref.~\cite{kas21}. 
The superscript $q$ denotes 
$\nu$ (neutron, $ t_z= 1/2$) or $\pi$ (proton, $t_z =-1/2$). 
The average particle number is fixed at the desired value by adjusting the chemical potential $\lambda^q$. 
When neutrons and/or protons are unpaired, the chemical potential is adapted to lie 
in the center of the highest occupied and the lowest unoccupied orbitals. 
Assuming the system is axially symmetric, 
the KSB equation (\ref{HFB_eq}) is block diagonalized 
according to the quantum number $\Omega$, the $z$-component of the angular momentum. 

The excited states $| i \rangle$ are described as 
one-phonon excitations built on the ground state $|0\rangle$ of the mother nucleus as 
\begin{align}
| i \rangle &= \hat{\Gamma}^\dagger_i |0 \rangle, \\
\hat{\Gamma}^\dagger_i &= \sum_{\alpha \beta}\left\{
X_{\alpha \beta}^i \hat{a}^\dagger_{\alpha}\hat{a}^\dagger_{\beta}
-Y_{\alpha \beta}^i \hat{a}_{\bar{\beta}}\hat{a}_{\bar{\alpha}}\right\},
\end{align}
where $\hat{a}^\dagger$ and $\hat{a}$ are 
the quasiparticle (qp) creation and annihilation operators that 
are defined in terms of the solutions of the KSB equation (\ref{HFB_eq}) with the Bogoliubov transformation.
The phonon states, the amplitudes $X^i, Y^i$ and the vibrational frequency $\omega_i$, 
are obtained in the quasiparticle-random-phase approximation (QRPA): the linearized time-dependent density-functional theory for superfluid systems~\cite{nak16}. 
The EDF gives the residual interactions entering into the QRPA equation. 
For the axially symmetric nuclei, the QRPA equation 
is block diagonalized according to the quantum number $K=\Omega_\alpha + \Omega_\beta$. 
The deformed QRPA developed in Refs.~\cite{yos08,yos13b} is extended to describe 
the response to the spin-dependent operators.

\subsection{Numerical procedures}

To describe the developed neutron skin and the neutrons pair correlation 
coupled with the continuum states that emerge uniquely in neutron-rich nuclei, 
I solve the KSB equation in the coordinate space using cylindrical coordinates
$\boldsymbol{r}=(\rho,z,\phi)$ with a mesh size of
$\Delta\rho=\Delta z=0.6$ fm and a box
boundary condition at $(\rho_{\mathrm{max}},z_{\mathrm{max}})=(14.7, 14.4)$ fm. 
Since I assume further the reflection symmetry, only the region of $z\geq 0$ is considered. 
The qp states are truncated according to the qp 
energy cutoff at 70 MeV, and 
the qp states up to the magnetic quantum number $\Omega=23/2$
with positive and negative parities are included. 
I introduce the truncation for the two-quasiparticle (2qp) configurations in the QRPA calculations,
in terms of the 2qp-energy as 60 MeV. 

For the normal (particle--hole, p--h) part of the EDF,
I mainly employ the SGII functional~\cite{gia81}. 
To complement the discussion, I use the SkP functional~\cite{dob84} which has a different spin-isospin property to SGII.
For the pairing energy, I adopt the one in Ref.~\cite{yam09}
that depends on both
the IS and IV densities, 
in addition to the pair density, with the parameters given in
Table~III of Ref.~\cite{yam09}. 
The same pairing EDF is employed for the $^1S$ pn-pairing 
in the proton--neutron QRPA (pnQRPA) calculation, 
while the linear term in the IV density is dropped. 
The effect of the dynamic IS ($T=0$) paring is investigated by changing the strength of the interaction. 
Note that the pnQRPA calculations including the dynamic IS pairing with 
more or less the same strength as the $^1S$ pairing 
describe well the characteristic low-lying GT
strength distributions in the light $N \simeq Z$ nuclei~\cite{fuj14,fuj15,fuj19}.

\section{Results and discussion}\label{result}

I consider the response of the Ca isotopes with the mass number $A=42\text{--}70$ and 
the Ni isotopes with $A=52\text{--}78$ to 
the IV spin-flip operator defined by
\be
\vec{F}_K
=\mu\dfrac{1}{\sqrt{2}} \sum_{ss^\prime}\sum_{tt^\prime}\int \dd \br
\psi^\dagger(\br s^\prime t^\prime)\psi(\br s t)
\langle s^\prime|\sigma_K|s\rangle\langle t^\prime|\vec{\tau}|t \rangle, \label{IVspin_op}
\ee
where $\mu=\mu_{\rm N}(\mathrm{g}_{\nu}-\mathrm{g}_{\pi})$, with $\mu_{\rm N}=e\hbar/2mc$ being the nuclear magneton, 
$\mathrm{g}_{\nu}$, $\mathrm{g}_{\pi}$ the spin g factor of a neutron and a proton, and $\sigma_K$ and 
$\vec{\tau}=(\tau_{+1},\tau_0,\tau_{-1})$ denote the spherical components of the Pauli matrix of spin and isospin. 
With this operator, the spin modes of the neutral (isospin unchanging: $\tau_0$) and isospin changing ($\tau_{\pm 1}$) excitations 
are investigated in a unified way. 
Note that I use here a probing operator that is $\mu \sqrt{2\pi}$ times the one used in Ref.~\cite{aue84}.
The IV spin-M1 and GT operators are indeed related to the IV spin-flip operator (\ref{IVspin_op}) 
as $F^{\textrm{IV,s}}_K(\textrm{M1})=\sqrt{\frac{3}{8\pi}}F_{K,0}$ and 
$\vec{F}_K(\textrm{GT})=\frac{g_A}{\mu}\vec{F}_K$ when the contribution of the meson-exchange and isobar currents is discarded~\cite{ric90,fuj11}.
Here $g_A$ is the axial-vector coupling constant. 

The Ca and Ni isotopes are all spherical in the ground state within the present model. 
The neutrons are paired in all the isotopes except $^{70}$Ca, $^{56}$Ni, and $^{78}$Ni 
while the protons are unpaired except in ${}^{52}$Ni. 

\subsection{Neutral channel: IV spin-M1 states in Ca isotopes}\label{result_M1}
According to the shell model, 
in a spin-saturated closed-shell nucleus, such as $^{40}$Ca, 
the M1 excitations are forbidden to occur. 
Thus the spin-flip excitations in the Ca isotopes that I investigate here are by virtue of the excitation 
of neutrons. 
It is noticed, however, that the M1 strengths in $^{40}$Ca were observed around 10 MeV~\cite{pri82} and 
the beyond-RPA correlation is indispensable to describe such a forbidden state~\cite{kam89}. 

\begin{figure}[t]
\begin{center}
\includegraphics[scale=0.39]{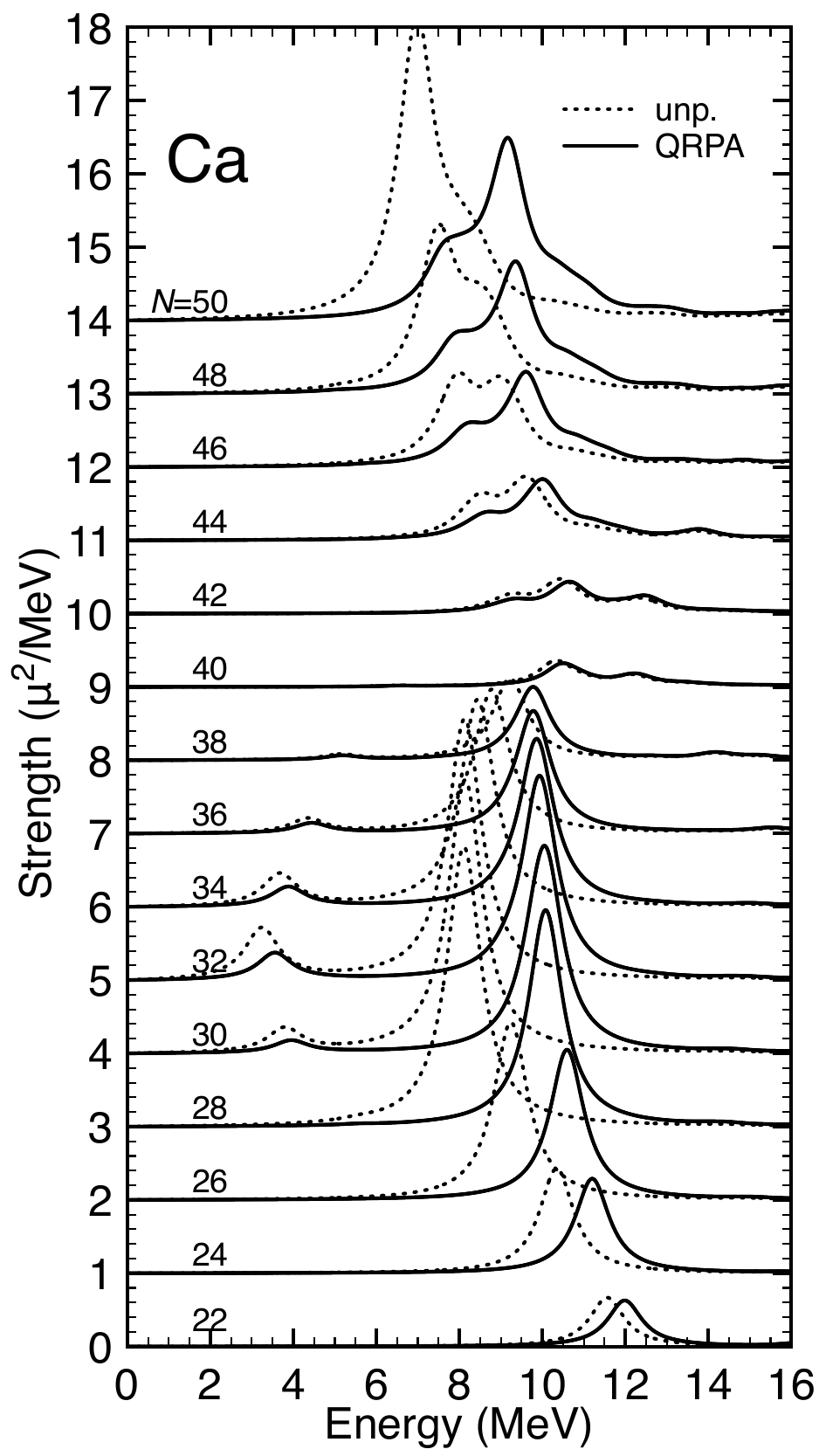}
\caption{\label{fig:Ca_spin_strength} 
Calculated distributions (shifted) of the IV spin-flip 
transition strengths in the neutral channel 
as functions of the excitation energy by employing the SGII functional. 
The smearing parameter $\gamma=1$ MeV is used. 
The results obtained without the RPA correlations (unp.) are depicted by the dotted lines. 
}
\end{center}
\end{figure}

\begin{figure}[t]
\begin{center}
\includegraphics[scale=0.4]{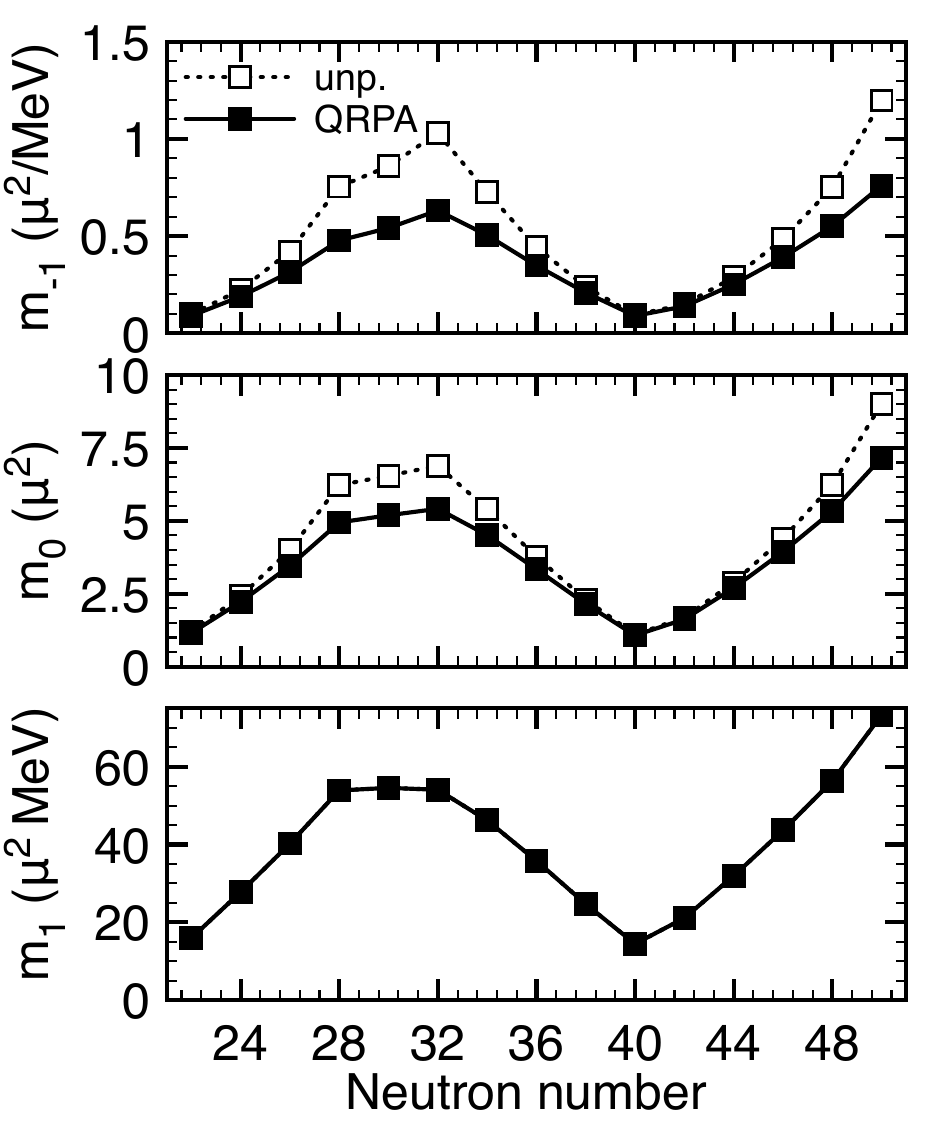}
\caption{\label{fig:Ca_spin_sum} 
Moments $m_k$ ($k=-1, 0$ and $1$) of the transition strengths in the Ca isotopes as functions 
of the neutron number. 
The results obtained without the RPA correlations (unp.) are depicted by the dotted lines with open symbols. }
\end{center}
\end{figure}

Figure~\ref{fig:Ca_spin_strength} shows the 
calculated transition strengths for the IV spin-flip excitation in the neutral channel, 
corresponding to the IV spin-M1 excitation: 
\begin{align}
S(E)&=\sum_{K=-1}^{1} \dfrac{\dd B(E,F_K)}{\dd E}, \label{eq:strength1}\\
\dfrac{\dd B(E,F_K)}{\dd E} &=\dfrac{2E\gamma}{\pi}\sum_{i}
\dfrac{\tilde{E}_{i} |\langle i|F_{K,0}|0 \rangle|^{2}}{(E^{2}-\tilde{E}_{i}^{2})^{2}+E^{2}\gamma^{2}}, \label{eq:strength}
\end{align}
where $\tilde{E}_{i}^{2}=(\hbar \omega_{i})^{2}+\gamma^{2}/4$~\cite{BM2}. 
The smearing width $\gamma$ is set to 1 MeV, which is supposed to simulate
the spreading effect, $\Gamma^\downarrow$, missing in the QRPA.

In the ${}^{42\text{--}48}$Ca isotopes, 
the spin-M1 resonant state is generated by 
the $\nu f_{7/2} \to \nu f_{5/2}$ excitation. 
The unperturbed state is shifted up in energy 
due to the residual interaction of the $\sigma_1\cdot\sigma_2 \tau_1\cdot\tau_2$ type.
To discuss the isotopic dependence of the strength distributions quantitatively, 
I show in Fig.~\ref{fig:Ca_spin_sum} the moment of the strengths defined by
\be
m_k=\sum_K \sum_i(\hbar \omega_i)^k|\langle i| F_{K,0}|0\rangle|^2
\ee 
for $k=-1, 0$, and 1. 
A linear rise seen in $m_0$ from $^{42}$Ca to $^{48}$Ca is 
understood by the occupation of neutrons in the $f_{7/2}$ orbital. 
A p--h-type transition strength is proportional to $u^2v^2$ of the 
Bardeen--Cooper--Schrieffer (BCS) amplitude. 
Since the $f_{5/2}$ orbital is completely empty $u_{f_{5/2}}=1$, 
the transition strengths develop monotonically as $v_{f_{7/2}}^2$ increases. 
Here, the occupation probability 
$v^2_{\alpha}$ can be evaluated by $\int \dd\br\sum_s |\varphi_{2,\alpha}(\br s)|^2$ using the qp wave functions. 
Comparing the results with and those without the RPA correlations, 
the effect of RPA correlations is also enhanced from $^{42}$Ca to $^{48}$Ca. 
The mean excitation energy evaluated by $m_1/m_0$ with (without) the RPA correlations is
13.8 (13.2), 12.5 (11.5), 11.6 (10.1), and 10.9 (8.7) MeV for $^{42,44,46,48}$Ca, respectively. 
This is again due to a gradual occupation of neutrons in the $f_{7/2}$ orbital. 
A two-body matrix element entering the QRPA equation is approximately proportional 
to $uvuv$ of the BCS amplitude. 
In the present case, only the $\nu f_{7/2} \to \nu f_{5/2}$ excitation 
accounts for the appearance of spin-M1 resonance in essence. 
Thus, the energy shift is due to the diagonal matrix element, which is 
proportional to $u^2_{f_{5/2}}v^2_{f_{7/2}}$ and evidently the occupation number of neutrons 
in the $f_{7/2}$ orbital. 

Experimentally, the $1^+$ state at 11.2 MeV and 10.3 MeV in $^{42}$Ca and $^{48}$Ca 
was populated by the inelastic electron scattering~\cite{ste80}. 
In the present model, the IV-spin M1 strength is concentrated on 
the $1^+$ state at 11.2 MeV and 10.1 MeV in $^{42}$Ca and $^{48}$Ca. 
A good agreement with the experiment shows the validity of the theoretical calculations. 

One sees an onset of the low-energy state in $^{50\text{--}56}$Ca. 
This is due to the $\nu p_{3/2} \to \nu p_{1/2}$ excitation. 
The increase in $m_0$ from $^{50}$Ca to $^{52}$Ca is understood 
by the occupation of neutrons in the $p_{3/2}$ orbital similarly in the case of the $f_{7/2}$ orbital in $^{42\text{--}48}$Ca. 
When the neutron number increases further, 
neutrons start to occupy the $p_{1/2}$ orbital. 
Then, the occupation of neutrons in the $p_{1/2}$ orbital 
diminishes the transition strength as $u^2_{p_{1/2}}v^2_{p_{3/2}}=1-v^2_{p_{1/2}}$.
A linear reduction in $m_0$ and the energy shift due to the RPA correlations 
is seen up to $^{60}$Ca, where the occupation of neutrons in the $f_{5/2}$ orbital is fulfilled. 
Note that the $f_{5/2}$ orbital is bound though the separation energy $E_{f_{5/2}}-\lambda^\nu$ is as low as 1.4 MeV. 
In $^{60}$Ca, the spin excitation is forbidden. 
A tiny strength, however, appears in the calculation. 
The appearance of the strengths is due to the pairing of neutrons. 
In the present calculation, the occupation probability of the $\nu 1g_{9/2}$ orbital is 0.08.

Beyond $^{60}$Ca, a linear increase in $m_0$ and the energy shift due to the RPA correlations 
is mainly due to a gradual occupation of neutrons in the $1g_{9/2}$ orbital. 
A distinct feature in the neuron-rich nuclei beyond $N=40$ is that 
the $g_{7/2}$ orbital is no more a bound state. 
Furthermore, there can be an appreciable contribution of 2qp excitations in the continuum 
because the resonance shape is not simply a single peak as calculated below $N=40$. 
Therefore, the spin-M1 state acquires a wider width due to a stronger coupling to the continuum states. 
Since the continuum states are discretized in the present calculation, 
a quantitative discussion on the resonance shape is not easy. 
It is thus an interacting future work to extend the continuum calculation like in Ref.~\cite{ham99} to heavier nuclei 
and discuss the neutron-number dependence of the width of the spin-M1 resonance~\cite{kam93}. 
The evolution of the spin-M1 strength distributions in the Ca isotopes is similar to 
the one predicted by the relativistic EDF approach~\cite{ois20}. 
It is noticed that the pairing governing the isotopic dependence of the RPA correlations in the spin-M1 excitation 
is the static pairing effect. 
The dynamic pairing caused by the residual pair interaction 
does not affect the collectivity of the unnatural parity states in the present model, 
while it enhances the collectivity of the natural parity states as discussed, {\it e.g.}, in Refs.~\cite{mat01,mat04,yos06b,yos06,yos08b,yos21}.

The inverse energy-weighted sum $m_{-1}$ shown in Fig.~\ref{fig:Ca_spin_sum}
is related to the IV spin susceptibility~\cite{lip89} as
\be
\chi_z \equiv 2\mu^2\sum_{i}\dfrac{|\langle i|S_z T_z|0\rangle |^2}{\hbar\omega_i}=\dfrac{m_{-1}}{3},
\ee
where $S_zT_z=
\int \dd\br\sum \psi^\dagger(\br s^\prime t^\prime)\psi(\br s t)\langle s^\prime|\frac{\sigma_z}{2}|s\rangle
\langle t^\prime|\tau_0|t \rangle$. 
This is of particular interest because it is not sensitive to the high-energy side of the response 
and hence is less affected by the two-particle--two-hole (2p2h) excitations~\cite{eri84}.
For nuclear matter, the IV magnetic susceptibility per nucleon is simply expressed~\cite{alb82} as
\be
\bar{\chi}_z=\dfrac{1+\frac{1}{3}F_1}{1+G_0^\prime}\mu^2 N_0
\label{eq:susc}
\ee
with the Landau parameters $N_0$, $G_0^\prime$ and $F_1$, and 
$\mu^2 N_0$ is regarded as 
the Pauli paramagnetic susceptibility. 
The susceptibility of the correlated systems is suppressed 
due to the repulsive nature of the interaction $G_0^\prime >0$. 
Note that the unperturbed value displayed in the top panel of Fig.~\ref{fig:Ca_spin_sum} 
includes the effect of $F_1$, {\it i.e.}, the effective mass that reduces the susceptibility 
and the like-particle pairing that also reduces the susceptibility as for the Belyaev inertia~\cite{bel65}. 
In the present case for the Ca isotopes, 
since protons do not take part in generating the collectivity, 
the spin-M1 state is simply described as a neutron p--h excitation of the 
spin--orbit partners. 
Therefore, the reduction of the susceptibility strongly depends on the 
shell structure and the pairing correlation of neutrons. 
The effect of the residual interaction is seen more clearly than in $m_0$.

\subsection{Charge-exchange channel: GT states in Ca isotopes}\label{result_GT}

\begin{figure}[t]
\begin{center}
\includegraphics[scale=0.39]{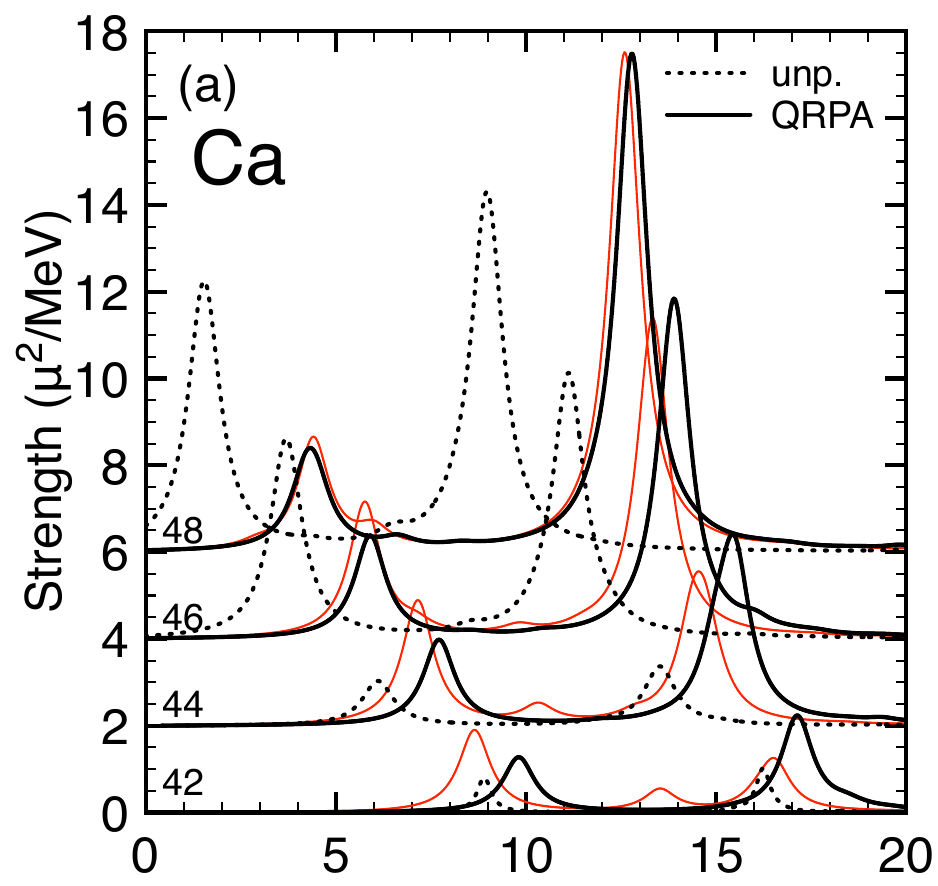}
\includegraphics[scale=0.39]{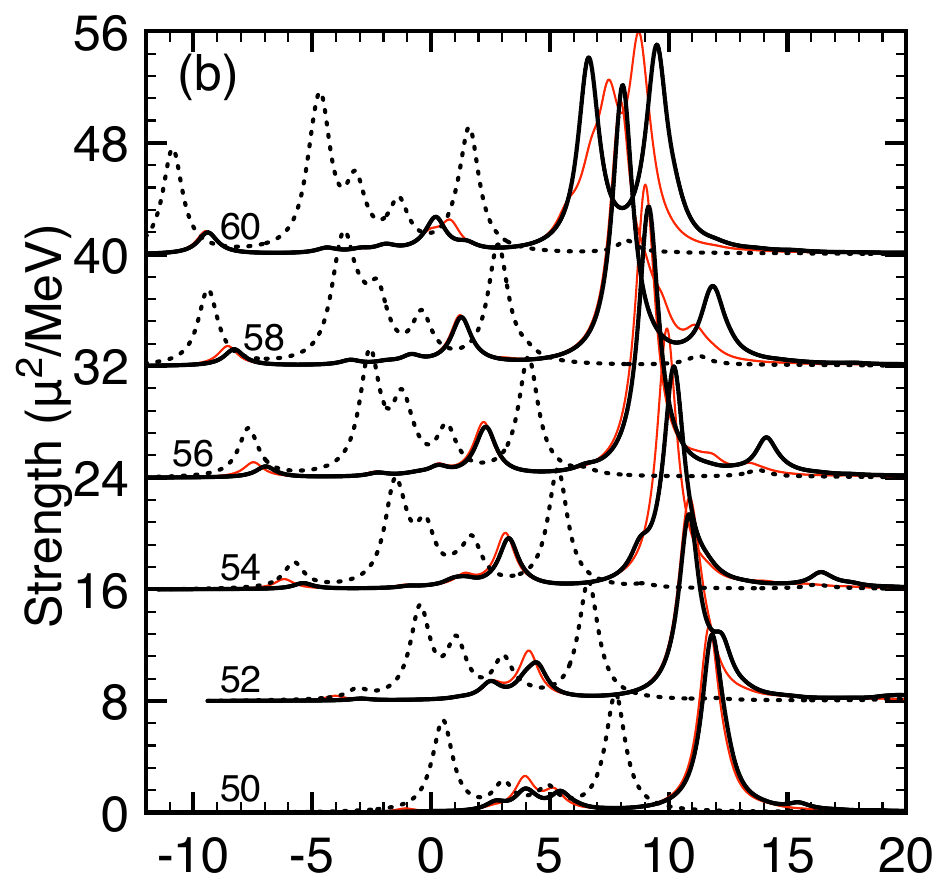}
\includegraphics[scale=0.39]{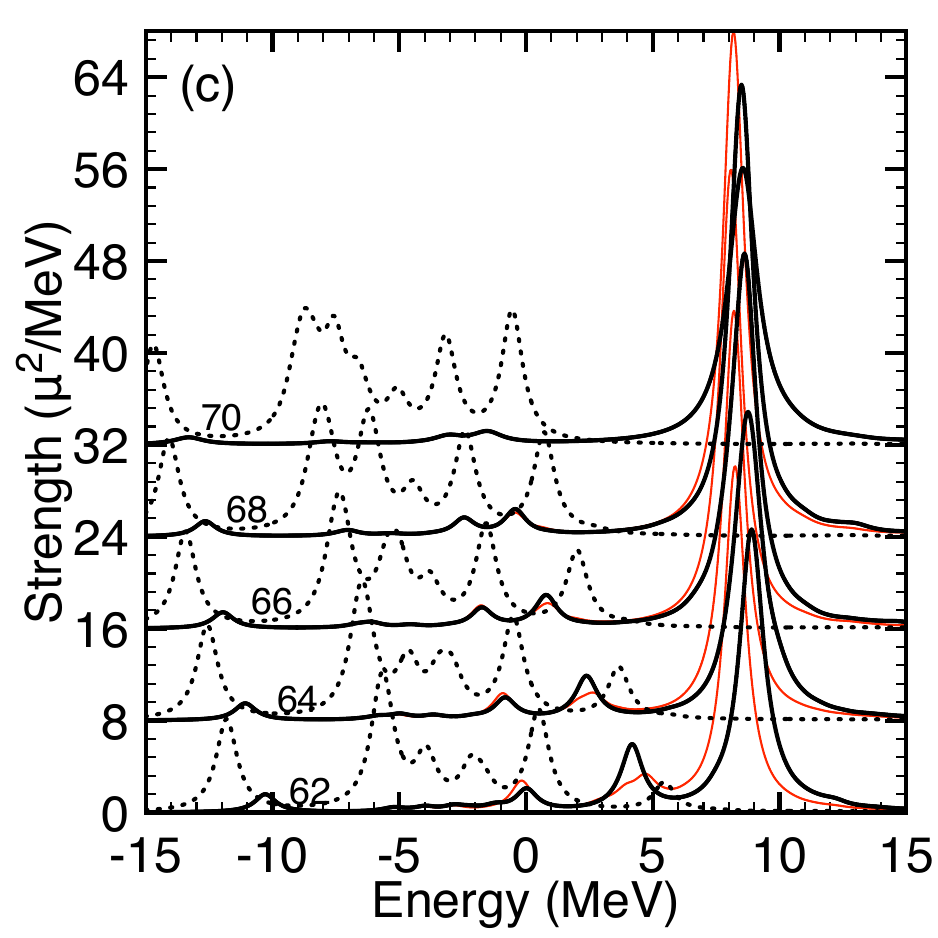}
\caption{\label{fig:Ca_GT} 
Similar to Fig.~\ref{fig:Ca_spin_strength} but for the response to the $t^{-}$ channel of the operator (\ref{IVspin_op}) 
in (a) $^{42\text{--}48}$Ca ($f_{7/2}$ shell), (b) $^{50\text{--}60}$Ca ($p\mathchar`-f_{5/2}$ shell) 
and (c) $^{62\text{--}70}$Ca ($g_{9/2}$ shell). The results including the $T=0$ (spin-triplet) pairing 
are drawn by the thin lines. 
}
\end{center}
\end{figure}

Figure~\ref{fig:Ca_GT} shows the 
calculated transition strengths for the IV spin-flip excitation in the charge-exchange channel ($t_-$), 
corresponding to the GT excitation, as given similarly to Eqs.~(\ref{eq:strength1}), (\ref{eq:strength}) 
except by replacing the excitation energy $E$ with $E_T=E-\lambda^\nu+\lambda^\pi$ 
so that the excitation is with respect to the ground state of the mother (target) nucleus. 
The transition in the $t_+$ channel is strongly suppressed as 
the excitation of protons in the neutral channel. 

Let me discuss the GT excitations in $^{42\text{--}48}$Ca shown in Fig.~\ref{fig:Ca_GT}(a). 
One sees a two-hump structure while there is only a single peak in the spin-M1 response. 
These states are mainly constructed by the $\nu f_{7/2}\to\pi f_{7/2}$ and $\nu f_{7/2}\to\pi f_{5/2}$ configurations.
Since the $(\nu f_{7/2})^2$ configuration couples to $S=0$, this does not take part in the 
formation of the spin-M1 state. 
With the growth of the neutron number, the transition strength increases. 
This is understood by a gradual increase in the occupation of neutrons in the $f_{7/2}$ orbital 
as discussed for a linear rise in the transition strengths in the neutral channel.
The excitation energy 
is predominantly given by the unperturbed energy and the diagonal matrix element of the 
$\sigma_1\cdot\sigma_2 \tau_1\cdot\tau_2$ interaction, and 
the two peaks are shifted up in energy due to a repulsive character 
of the interaction. The energy shift is also nearly proportional to the 
occupied neutron number in the $f_{7/2}$ orbital.

The IS pairing affects the GT states~\cite{eng99,yos13} and 
explains well 
the enhancement of the strength for the low-energy GT state in $^{18}$O ($^{42}$Ca)~\cite{fuj14,fuj15,fuj19,bai14,sag15}, 
where the particle--particle (p--p) type excitations of $\nu p_{3/2}\to \pi p_{3/2}$ ($\nu f_{7/2}\to \pi f_{7/2}$)
and $\nu p_{3/2}\to \pi p_{1/2}$ ($\nu f_{7/2}\to \pi f_{5/2}$) 
cooperatively participate in generating the 
low-energy GT state. 
The numerical results including the dynamic IS pairing are 
depicted by the thin line in Fig.~\ref{fig:Ca_GT}. 
The effect of the IS pairing is strong in $^{42}$Ca as discussed in Ref.~\cite{bai14}. 
Here, the strength of the IS pair interaction was set as the same as that 
for the IV pair interaction. 

Experimentally, the GT strength is concentrated on the $1^+$ state at 0.61 MeV
in $^{42}$Sc, that is observed by the $^{42}$Ca($^3$He,$t$) reaction~\cite{fuj14,fuj15}. 
In the present model, the $1^+$ state appears at 1.79 MeV above the isobaric analog state (IAS) (the ground state of $^{42}$Sc).
With the IS pair interaction being included, I obtain the $1^+$ state at 0.63 MeV. 
In the $^{44}$Ca($^{3}$He,$t$)$^{44}$Sc reaction, the $1^+$ state was observed at 2.11 MeV below the IAS~\cite{fuj13}. 
The calculation gives the $1^+$ state at 0.38 MeV above and 0.17 MeV below the IAS without and with the IS pair interaction, respectively. 
The IS pairing is still active, but the effect is weaker than in $^{42}$Ca. 
In the $^{48}$Ca($^{3}$He,$t$)$^{48}$Sc reaction, the $1^+$ state was observed at 4.18 MeV below the IAS~\cite{gre07}. 
The calculation gives several $1^+$ states in low energies: 2.4--2.6 MeV below the IAS.

As neutrons occupy the $2p$ orbitals, 
the $\nu p_{3/2, 1/2}\to \pi p_{3/2,1/2}$ excitations 
show up in low energies as depicted by the dotted lines in Fig.~\ref{fig:Ca_GT}(b). 
Thanks to the RPA correlations, most of the strengths are 
gathered into the giant resonance though there remains a tiny strength in the low energy region. 
These low-lying states are responsible for the $\beta$ decay. 

\begin{figure*}[t]
\begin{center}
\includegraphics[scale=0.4]{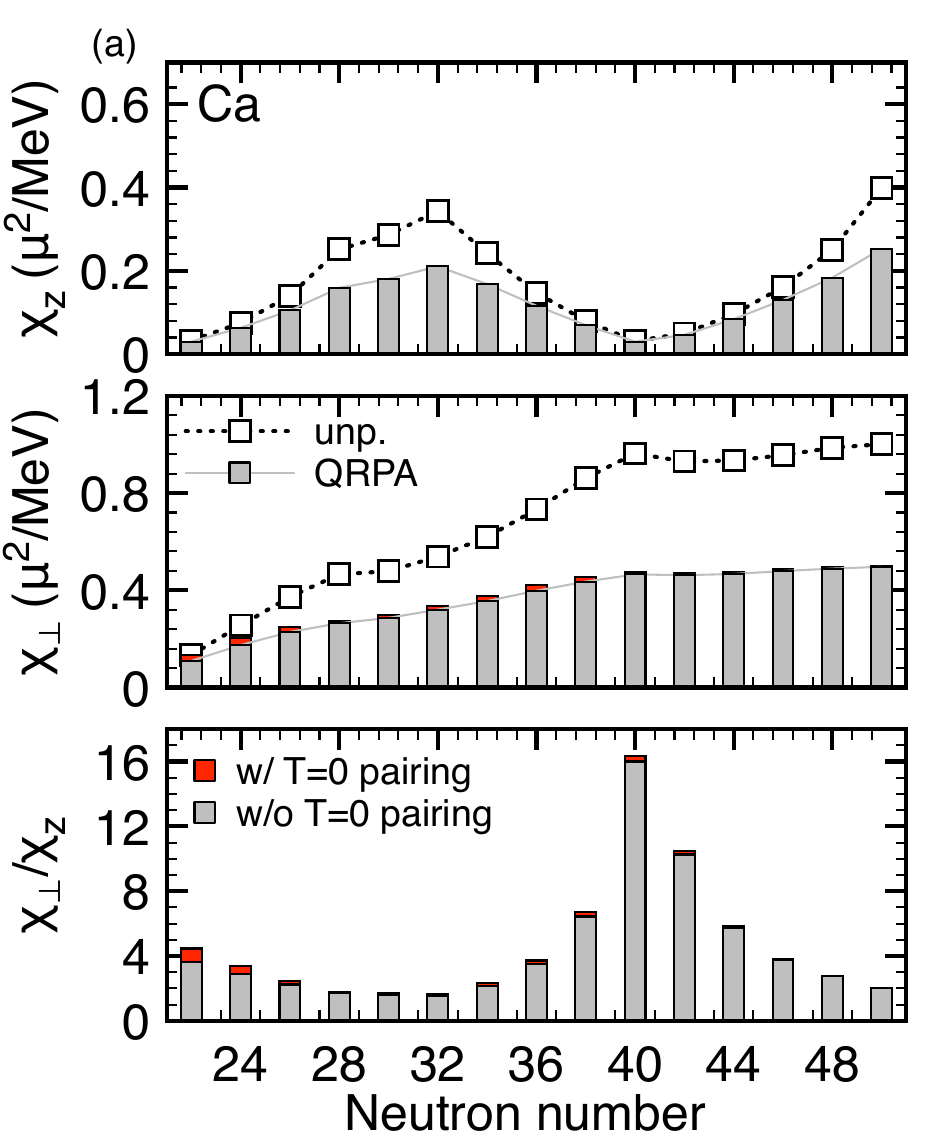}
\includegraphics[scale=0.4]{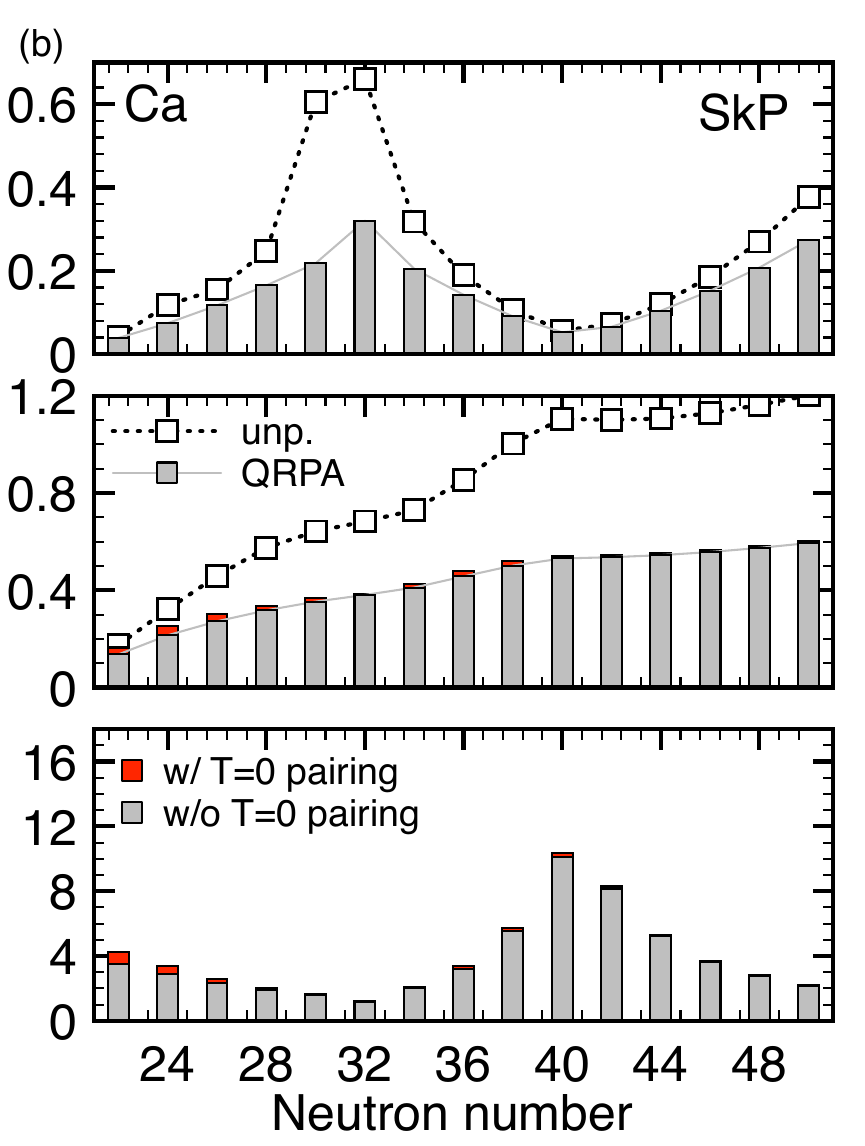}
\caption{\label{fig:Ca_GT_sum} 
Results obtained by using the (a) SGII functional and (b) SkP functional. 
Top: IV spin susceptibility $\chi_{z}$ of the Ca isotopes obtained 
with and without the RPA correlations.
Middle: IV spin susceptibility $\chi_\perp$ in the charge-exchange channel. 
Bottom: Ratio of the IV-spin susceptibilities $\chi_\perp/\chi_z$ obtained 
with and without the dynamic $T=0$ (spin-triplet) pairing. 
}
\end{center}
\end{figure*}

Beyond $N=34$, one sees an appreciable amount of the unperturbed strengths is developed in low energies 
with an increase in the neutron number. 
This comes from the $\nu f_{5/2}\to \pi f_{7/2}$ excitation. 
Due to the RPA correlations, however, most of the strengths 
are brought together into the high-lying giant resonance. 
The remaining low-lying state is expected to be affected by the IS pairing because of 
the p--p nature of the configuration. 
In $^{54,56}$Ca, the occupation probability of neutrons in the $f_{5/2}$ orbital is 0.2 and 0.4.
In the present cases, the low-lying state is lowered in energy and enhanced in strengths only slightly, 
as shown by the thick line in Fig.~\ref{fig:Ca_GT}(b). 
This is because the $\nu f_{7/2}\to \pi f_{5/2}$ excitation is purely a p--h type and 
is not affected by the IS pairing.

The IS pairing affects the giant resonance, as neutrons partially occupy the $g_{9/2}$ orbital. 
Beyond $N=40$, 
the $\nu g_{9/2}\to \pi g_{9/2}$ configuration 
appears more clearly in the high energy region 
according to a gradual increase in the occupation probability of neutrons in the $g_{9/2}$ orbital. 
Since this configuration is a p--p type excitation, 
the IS pairing is active. 
As one can see in Fig.~\ref{fig:Ca_GT}(c), 
the giant resonance is lowered in energy due to 
the attractive nature of the IS pairing.

In neutron-rich isotopes, the number of the available p--h or 2qp excitations increases 
because of the imbalanced Fermi level of neutrons and protons. 
Therefore, the GT strengths are concentrated in the giant resonance in $^{62\text{--}70}$Ca 
as a collective effect 
notwithstanding that a significant fraction of the GT strengths 
is found inside the $Q_\beta$ window without the RPA correlations. 
The super-allowed GT resonance predicted in the light neutron drip-line nuclei~\cite{sag93} 
is thus unlikely to occur in the Ca isotopes even near the drip line.

To investigate the magnetic property systematically, I introduce 
the static susceptibility. 
The IV spin susceptibility in the charge-exchange channel 
is given as
\be
\chi_\perp = \mu^2\left[ 
\sum_{i}\dfrac{|\langle i|S_z T_-|0\rangle |^2}{\hbar\omega_i}+
\sum_{i}\dfrac{|\langle i|S_z T_+|0\rangle |^2}{\hbar\omega_i}
\right],
\ee
where $S_zT_{\pm}=
\int \dd\br\sum \psi^\dagger(\br s^\prime t^\prime)\psi(\br st)\langle s^\prime|\frac{\sigma_z}{2}|s\rangle
\langle t^\prime|\tau_{\pm 1}|t \rangle$. 
For nuclear matter, one may expect $\chi_\perp$ coincides with $\chi_z$. 
The IV spin susceptibility in the charge-exchange channel 
is reduced due to the 
RPA correlations as in the neutral channel.

I show in the middle panel of Fig.~\ref{fig:Ca_GT_sum}(a) the calculated IV spin susceptibilities $\chi_\perp$ obtained 
with and without the RPA correlations. 
As the neutron number increases, 
the p--h or 2qp excitations possessing the non-vanishing GT matrix-element appear in low energies.
The susceptibility in the free system thus increases. 
Since the operator (\ref{IVspin_op}) changes only the direction of spin and isospin, 
and does not change the spatial structure, 
the p--h or 2qp configurations in the $0\hbar \omega_0$ excitation are only possible 
to appear in the GT response. 
Around $N=40$, 
the $-1\hbar \omega_0$ excitations start to appear in low energies ~\cite{yos17}, 
and the $0\hbar \omega_0$ excitations show up in a relatively higher energy region. 
Therefore, $\chi_\perp$ keeps almost unchanged beyond $N=40$. 
The effect of the RPA correlations becomes apparent as the collectivity becomes strong. 
With an increase in the neutron number, the number of p--h or 2qp excitations 
in the $pf$-shell increases, and they make a coherent contribution 
to the formation of the giant resonance. 
One can thus see a systematic reduction of $\chi_\perp$ from the unperturbed one in the range of $22\leq N \leq 40$.

The dynamic IS pairing acts on the spin response oppositely to the p--h residual interaction, 
the former causes an attractive shift of the transitions and enhances the susceptibility 
while the latter a repulsive shift and reduces the susceptibility. 
For nuclear matter one may assume to express the ratio of the IV spin susceptibilities as 
\be
\dfrac{\chi_\perp}{\chi_z}=\dfrac{1+G_0^\prime}{1+G_0^\prime+V_{\textrm{pair}}}=1-\dfrac{V_{\textrm{pair}}}{1+G_0^\prime+V_{\textrm{pair}}}
\ee
with the IS pair interaction $V_{\textrm{pair}}<0$. 
For asymmetric systems, there should be a correction due to the neutron excess. 
A recently developed technique~\cite{dav19} can be used to 
evaluate this, but it is beyond the scope of the present work.
It is also noticed that the dynamic IS pairing affects the 
p--h type excitation only through the static IV pairing. 
Within the scope of consideration, roles of the IS pairing can thus be investigated only at a minimum. 
In plotting the ratio $\chi_\perp/\chi_z$, nevertheless, 
one can expect to see the magnetic properties from different perspectives.
Shown in the bottom panel of Fig.~\ref{fig:Ca_GT_sum}(a) is the ratio for the Ca isotopes. 
Since protons do not take part in the spin excitations as mentioned in the beginning, 
one sees a high asymmetry between $\chi_z$ and $\chi_\perp$ even in the case of $N \sim Z$. 
The enhancement in the ratio $\chi_\perp/\chi_z$ is governed by 
the shell effect in $\chi_z$, and displayed is only 
a tiny contribution of the dynamic IS pairing to the susceptibility.

Before investigating the Ni isotopes, I am going to discuss briefly how robust the prediction is.
Figure~\ref{fig:Ca_GT_sum}(b) shows the results obtained by employing the SkP functional. 
The residual interaction in the spin--isospin channel plays a significant role in the IV spin excitations. 
It is noted that the Landau parameter $G_0^{\prime}$ of the SGII and SkP functionals are 0.93 and 0.06, respectively~\cite{ben01}.
The isotopic evolution of the IV spin susceptibilities is essentially identical to those obtained using the SGII functional, 
while the susceptibilities are predicted to be larger. 
The enhancement of the susceptibilities is understood by Eq.~(\ref{eq:susc}); the SkP functional has 
a higher effective mass and a weaker residual interaction than SGII. 

\subsection{Ni isotopes}\label{Ni_isotopes}

\begin{figure}[t]
\begin{center}
\includegraphics[scale=0.4]{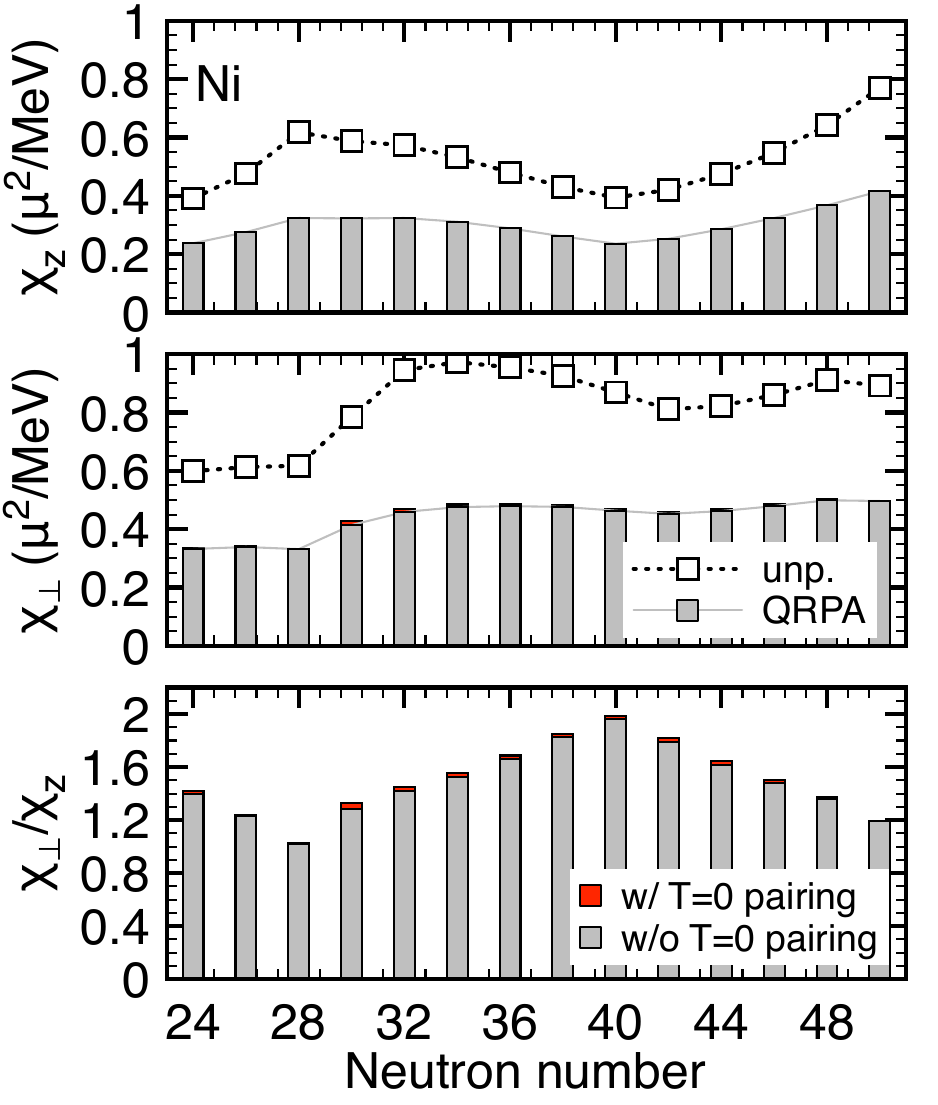}
\caption{\label{fig:Ni_sum} 
As Fig.~\ref{fig:Ca_GT_sum} but for the Ni isotopes. }
\end{center}
\end{figure}

\begin{figure}[t]
\begin{center}
\includegraphics[scale=0.39]{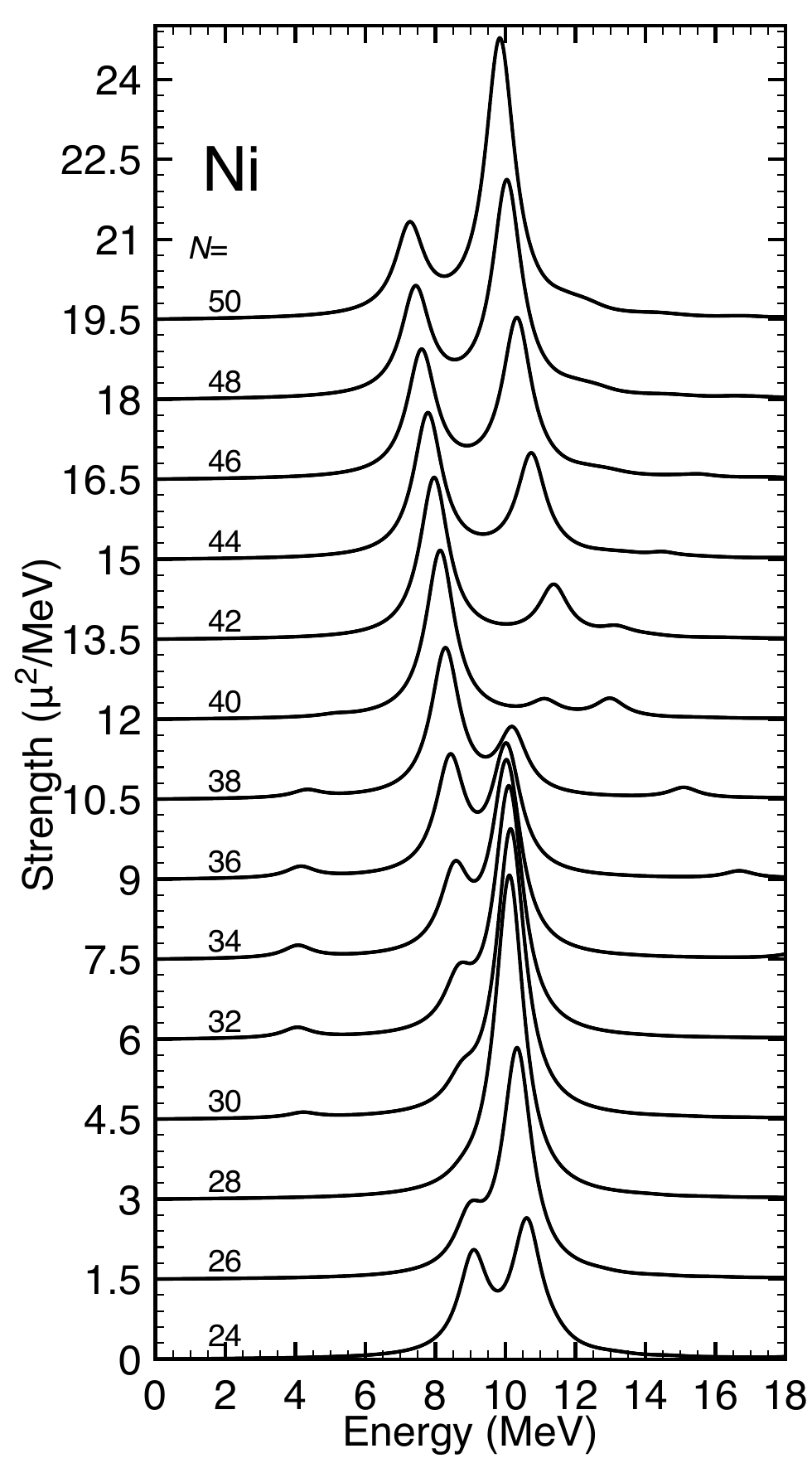}
\caption{\label{fig:Ni_spin_strength} 
As Fig.~\ref{fig:Ca_spin_strength} but for the Ni isotopes. 
}
\end{center}
\end{figure}

An outline of the isotopic dependence of the magnetic properties can be illustrated in 
the IV spin susceptibilities, $\chi_z$ and $\chi_\perp$. 
Thus I discuss the isotopic evolution of the collectivity in the spin responses 
for the Ni isotopes by showing 
Fig.~\ref{fig:Ni_sum}. 

Let me discuss first $\chi_z$. 
In the present case, the $\pi f_{7/2}\to\pi f_{5/2}$ excitation participates in the spin-M1 states in addition to 
the excitation of neutrons. 
As in the case for the Ca isotopes, 
the isotopic evolution of the collectivity is mainly explained by the 
occupation of neutrons in the $j_>$ orbital. 
To fortify a conclusion, I show in 
Fig.~\ref{fig:Ni_spin_strength} the strength distributions of the IV spin-M1 excitation. 
A gradual increase in the collectivity up to $N=28$, enhancement in the transition strength in the giant resonance region and 
suppression of the susceptibility due to RPA correlations, 
is because of the occupation of neutrons in the $f_{7/2}$ orbital. 
Taking a close look at the strength distribution, 
one sees that two peaks appear in the giant resonance region in $^{52}$Ni, 
while two peaks merge into a single peak in $^{56}$Ni. 
The lower-energy and higher-energy states are mainly generated by 
the $\pi f_{7/2}\to\pi f_{5/2}$ excitation 
and the $\nu f_{7/2}\to\nu f_{5/2}$ excitation, respectively. 
The energy difference for these p--h excitations is about 2 MeV in $^{52}$Ni, 
while about 0.2 MeV in $^{56}$Ni. 
Thus, the coherence between these excitations develops from $^{52}$Ni to $^{56}$Ni.
When neutrons occupy the $p_{3/2}$ orbital, a low-lying state shows up.
The appearance of the low-lying state has a contribution to the increase in $\chi_z$ in ${}^{50\text{--}52}$Ca. 
In $^{58\text{--}60}$Ni, however, the mixing between the 
$\pi f_{7/2}\to\pi f_{5/2}$ 
and $\nu f_{7/2}\to\nu f_{5/2}$ excitations becomes weak, 
which leads to the cancellation of $\chi_z$. 
As neutrons occupy the $p_{1/2}$ and $f_{5/2}$ orbitals, 
the collectivity decreases further. 
At $N=40$, the spin excitation is forbidden for neutrons. 
Beyond $N=40$, the $\nu g_{9/2} \to \nu g_{7/2}$ excitation starts to appear in the high energy. 
The lower-energy peak around 8--9 MeV and the higher-energy peak around 10--11 MeV 
are predominantly generated by the $\pi f_{7/2}\to\pi f_{5/2}$ excitation and the $\nu g_{9/2} \to \nu g_{7/2}$ excitation, 
respectively. 
As in the case for the Ca isotopes, 
the transition strengths in high energies develop with an increase in the occupation of neutrons in 
the $g_{9/2}$ orbital.

\begin{figure}[t]
\begin{center}
\includegraphics[scale=0.39]{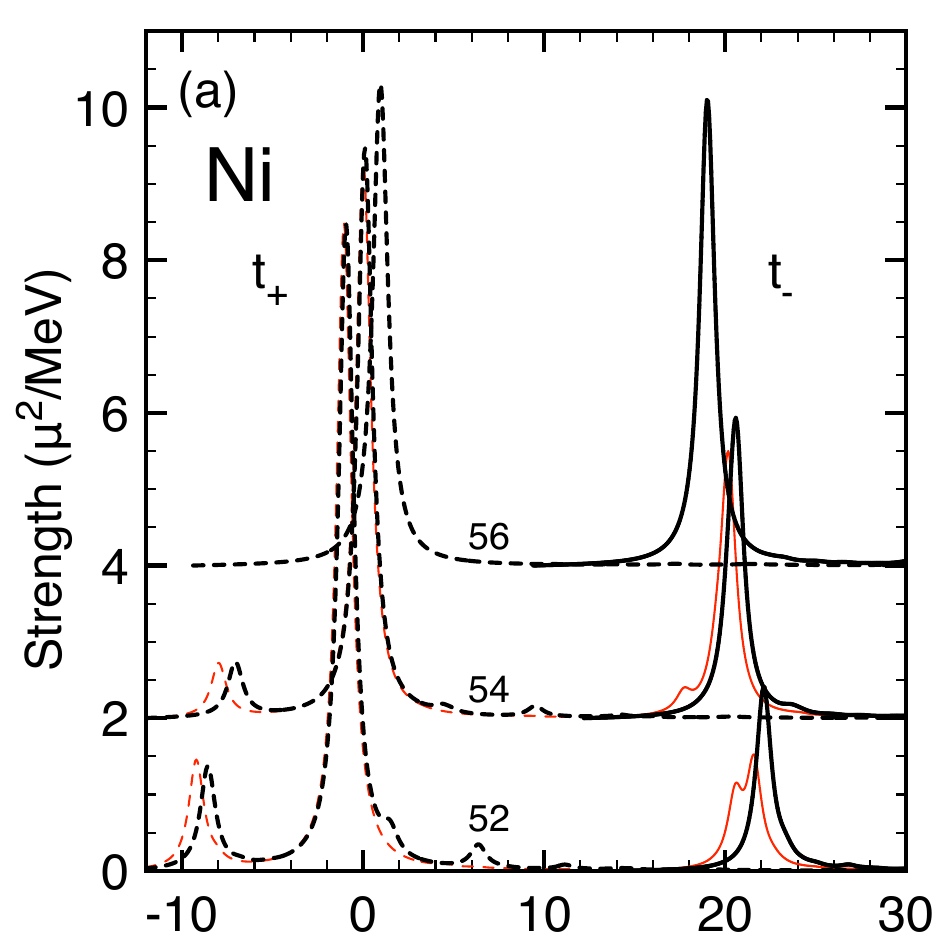}
\includegraphics[scale=0.39]{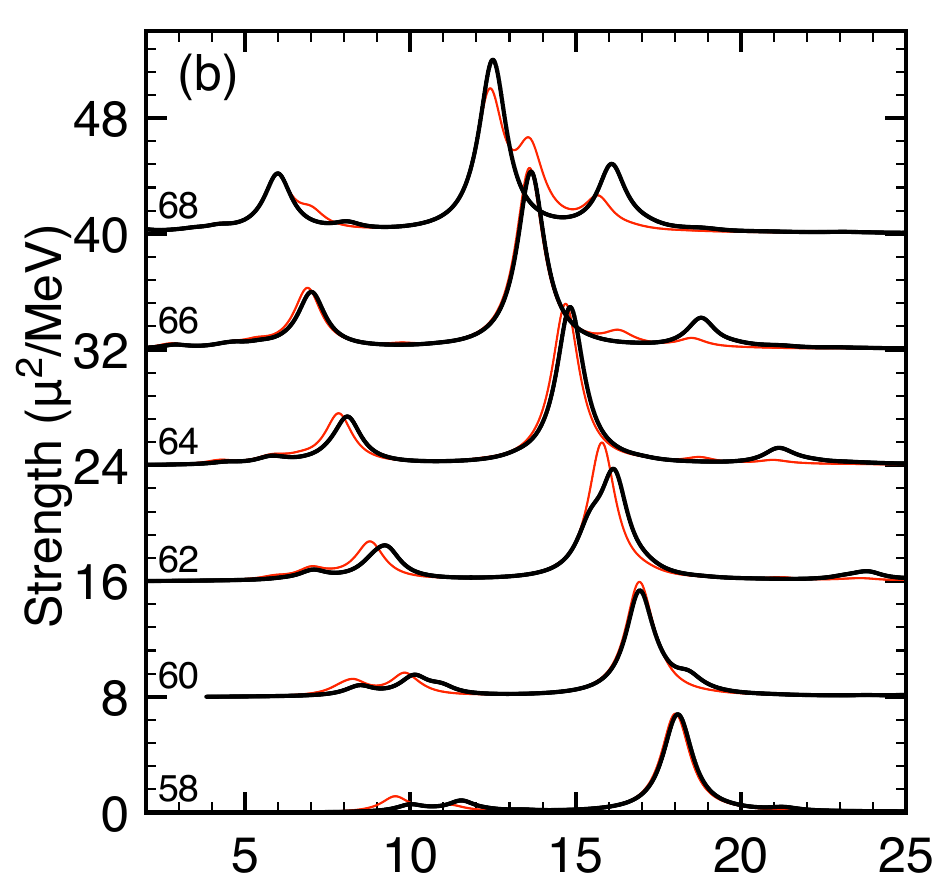}
\includegraphics[scale=0.39]{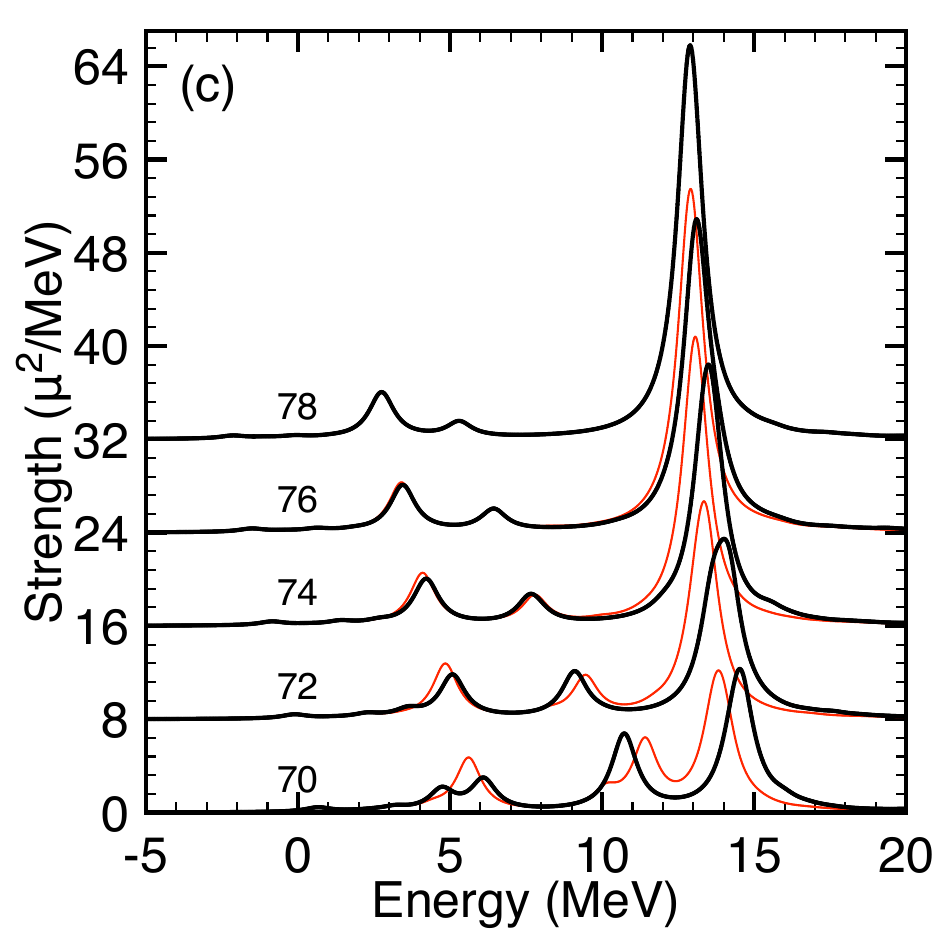}
\caption{\label{fig:Ni_GT} 
Similar to Fig.~\ref{fig:Ca_GT} but for 
(a) $^{52\text{--}56}$Ni ($f_{7/2}$ shell), (b) $^{58\text{--}68}$Ni ($p\mathchar`-f_{5/2}$ shell) 
and (c) $^{70\text{--}78}$Ni ($g_{9/2}$ shell). 
}
\end{center}
\end{figure}

I then discuss $\chi_\perp$. 
In $^{52\text{--}56}$Ni, the excitation in the $t_+$ channel has an appreciable 
contribution to the IV spin susceptibility as well as that in the $t_-$ channel. 
Therefore, $\chi_\perp$ value is higher than in the Ca isotopes below $N=28$. 
To see what is happening more clearly, I show 
in Fig.~\ref{fig:Ni_GT}(a) the transition strengths in the $t_+$ channel by 
the dashed line, where 
$E$ is replaced by $E_T=E+\lambda^\nu-\lambda^\pi$ in Eq.~(\ref{eq:strength}); 
the transition strengths in the $t_-$ channel are drawn by the solid line. 
The low-lying state in the $t_+$ channel 
is mainly generated by the $\pi f_{7/2}\to \nu f_{7/2}$ excitation. 
In $^{52}$Ni, the neutron occupation probability in the $f_{7/2}$ orbital is 0.48. 
This configuration is then considered as a hole--hole type excitation and is thus affected by the dynamic IS pairing, though not so strongly. 
The GT state in the $t_-$ channel is constructed by the 
$\nu f_{7/2}\to \pi f_{5/2}$ excitation, which can be considered as a p--p type excitation at $N=24$. 
Since the $\nu f_{7/2}\to \pi f_{7/2}$ excitation is forbidden, 
the collectivity and the influence by the IS pairing are weaker than in the Ca isotopes. 

Beyond $N=28$, one sees in the middle panel of Fig.~\ref{fig:Ni_sum} 
that the collectivity develops with an increase in the neutron number 
as in the Ca cases. 
At $N>34$, 
the collectivity does not increase in the Ni isotopes, 
while one has a gradual increase in the collectivity toward $N=40$ in the Ca isotopes. 
This different isotopic dependence results from the fact that 
the $\nu f_{5/2}\to \pi f_{7/2}$ excitation does not take part in generating the collectivity 
due to the Pauli effect in the Ni isotopes. 
Beyond $N=40$, the IV spin susceptibility keeps almost constant similarly in the Ca isotopes 
because 
the strength distribution in the low energy is almost unchanged and 
the $\nu g_{9/2}\to\pi g_{9/2}$ excitation appears in a high energy region. 
Note that the negative parity states show up instead in the low energy, leading to 
an interplay between the allowed and first-forbidden $\beta$-decays~\cite{yos19}. 
Since the $\nu g_{9/2}\to\pi g_{9/2}$ excitation is a p--p type excitation around $N=40$, 
the IS pairing affects the giant resonance as shown in Figs.~\ref{fig:Ni_GT}(b) and \ref{fig:Ni_GT}(c). 

Experimentally, the M1 resonance in $^{58,60,62}$Ni has been investigated by several approaches~\cite{lin76,met87,fuj07}, 
and the fragmentation of the strengths was found in the energy region of 8--15 MeV. 
Similarly, the GT strengths are fragmented in the low energy as well as in the resonance energy region~\cite{pop09,fuj07,fuj11}. 
Thus, it is not simple to investigate the individual states. 
Furthermore, the QRPA does not describe the spreading effect originating from the coupling to the 2p2h excitations. 
Therefore, the sum rule values or the spin susceptibilities are helpful to investigate the magnetic property systematically.

Finally, let me mention the isotopic dependence of the ratio of the IV spin susceptibilities, $\chi_\perp/\chi_z$, 
showing in the bottom panel of Fig.~\ref{fig:Ni_sum}.
In $^{56}$Ni, its value is unity, where protons and neutrons behave symmetrically. 
Furthermore, the IS pairing is ineffective for the spin susceptibility because of the shell closure. 
The collectivity becomes stronger above $N=28$ and stays almost constant beyond $N=34$ 
in the charge-exchange channel, 
while that in the neutral channel decreases toward $N=40$. 
Then, one sees a gradual increase in the ratio. 
Beyond $N=40$, the collectivity is enhanced in the neutral channel, 
leading to a decrease in the ratio as in the Ca isotopes. 
The effect of the IS pairing is much weaker than in the case of the Ca isotopes. 

\section{Summary}\label{summary}

A comparative study on the isotopic dependence of the magnetic properties 
has been performed by investigating the IV spin-flip excitations in the Ca and Ni isotopes. 
The responses in the neutral and charge-exchange channels were 
considered in a unified way. 
I made use of 
the nuclear EDF method for calculating the response functions 
based 
on the Skyrme--KSB and the QRPA. 
The like-particle QRPA and the proton--neutron QRPA were employed for the neutral 
and charge-exchange channels, respectively. 
The collective shift due to RPA correlations 
for the response in the neutral channel 
is mainly explained by the occupation probability 
of neutrons in the $j_>$ orbital, where the IV pairing of neutrons gives a dominant role. 
In the charge-exchange channel, 
many p--h or 2qp excitations have a coherent 
contribution to form a giant resonance in neutron-rich nuclei.
The dynamic IS pairing lowers the low-lying and giant resonance states sensitively to the shell structure. 
I have found that the isotopic evolution of the collectivity and the shell structure 
are nicely displayed by the IV spin susceptibility. 
A repulsive character of the residual interaction in the spin--isospin channel diminishes 
the susceptibility, while the IS pairing appearing in the charge-exchange channel opposes the suppression.

\begin{acknowledgments} 
This work was supported by the JSPS KAKENHI (Grants No. JP19K03824 and No. JP19K03872).
The numerical calculations were performed on Yukawa-21 
at the Yukawa Institute for Theoretical Physics, Kyoto University.
\end{acknowledgments}

\end{document}